\documentclass[letterpaper,10pt,final]{article}
\usepackage{amssymb, amsmath}
\usepackage{graphicx}

\title{Constraining a bulk viscous matter-dominated cosmological model using SNe Ia, CMB and LSS}
\author{Arturo Avelino, U. Nucamendi and F. S. Guzm\'an}
\date{\emph{\small Instituto de F\'{\i}sica y Matem\'aticas\\
Universidad Michoacana de San Nicol\'as de Hidalgo \\
Edificio C-3, A.P. 2-82, Ciudad Universitaria, CP. 58040\\
Morelia, Michoac\'an, M\'exico}}

\begin{document}

\maketitle

\begin{abstract}
We present and constrain a cosmological model which component is a 
pressureless fluid with bulk viscosity as an explanation for 
the present accelerated expansion of the universe. 
We study the particular model of a constant bulk viscosity 
coefficient $\zeta_{{\rm m}}$. The possible values of $\zeta_{{\rm m}}$ 
are constrained using the cosmological tests of SNe Ia 
Gold 2006 sample, the CMB shift parameter $R$ from the three-year WMAP
observations, the Baryon Acoustic Oscillation (BAO) peak $A$ from the Sloan Digital Sky
Survey (SDSS) and the Second Law of Thermodynamics (SLT).
It was found that this model is in agreement with the 
SLT using only the SNe Ia test. However when the model is 
submitted to the three cosmological tests together (SNe+CMB+BAO) 
the results are:  1.- the model violates the SLT, 2.- predicts a value of 
$H_0 \approx 53 \; {\rm km \cdot sec^{-1} \cdot Mpc^{-1}}$ for the 
Hubble constant, and 3.- we obtain a bad fit to data with a 
$\chi^2_{{\rm min}} \approx 400$ ($\chi^2_{{\rm d.o.f.}} \approx 2.2$). 
These results indicate that this model is ruled out by the observations.
\end{abstract}

We present a flat cosmological model which component is a 
pressureless fluid made of baryon and dark matter 
components with a constant bulk viscosity coefficient of the form
$\zeta_{{\rm m}}=$constant, which potentially could explain 
the present accelerated expansion of the universe. A bulk 
viscosity coefficient can produce a positive term in the 
second Friedmann equation that induces an acceleration  
\cite{Misner1973,Fabris2006} (i.e., $\ddot{a} \geq 0 \;$, with $a$ denoting
the scale factor and the dots 
derivatives with respect to the cosmic time). Similar models 
or analysis have been proposed also in 
\cite{Fabris2006,ArturoUlises2008,Colistete2007,XinHe2006a}.

		\paragraph{Model of the bulk viscous matter-dominated universe}
The energy-momentum tensor of a fluid composed by \emph{only} matter
(hereinafter, we call \emph{matter} to baryon and dark matter
components together) with bulk viscosity $\zeta_{{\rm m}}$ is
defined by $T_{\mu\nu}=\rho_{{\rm m}} u_\mu u_\nu
+ (g_{\mu\nu}+u_\mu u_\nu)P^*_{{\rm m}}$, where $P^*_{{\rm m}} = P_{{\rm m}}-
3\zeta_{{\rm  m}} \left( \dot{a}/a \right)$ (see e.g. \cite{Misner1973}). 
Here $P_{{\rm m}}$ is the pressure of the matter fluid. 
However, assuming pressureless matter: 
$P_{{\rm m}}=0 \; \Rightarrow \; P^*_{{\rm m}} =-3\zeta_{{\rm  m}} \left( \dot{a}/a 
\right)$. The subscript ``m'' stands for \emph{matter}.
The four-velocity vector $u_\mu$ is that of a comoving observer 
that measures the pressure $P^*_{{\rm m}}$ and the density $\rho_{{\rm m}}$ of the
matter fluid. 
The conservation of energy has the form 
$ a ( d \rho_{{\rm m}}/da) = 3\left( 3\zeta_{{\rm m}} H -\rho_{{\rm
m}} \right)$, where $H \equiv \dot{a}/a \;$ is the Hubble parameter. 
This equation in terms of the redshift reads 
$(1+z) \left(d\hat{\Omega}_{{\rm m}}/dz \right) + 
\hat{\Omega}^{1/2}_{{\rm m}} \; \tilde{\zeta}_0 - 3 \hat{\Omega}_{{\rm m}}=0$, 
where we have defined \footnote{We use this particular definition of $\hat{\Omega}_{{\rm m}}$ for computational conveniences.
$\rho^0_{\rm{critic}}\equiv 3H^2_0/8\pi G$ is the \emph{critical density} today, where $H_0$ is the Hubble constant.} 
 $\hat{\Omega}_{{\rm m}}(z) \equiv \rho_{{\rm m}} /\rho^0_{\rm{crit}}$, 
and the  dimensionless coefficient $\tilde{\zeta}_0  \equiv \zeta_{{\rm m}} (24\pi G/H_0)$. 
The exact solution to this equation is $\hat{\Omega}_{{\rm m}}(z)=(1/9)\left[ (\tilde{\zeta}_0 -3)(1+z)^{3/2}  - 
\tilde{\zeta}_0 \right]^2$.
On the other hand, a spatially flat geometry is assumed 
($k=0$) for the FRW metric cosmology as suggested by WMAP.
Thus, the first Friedmann equation for a flat universe composed by bulk viscous matter has the form:
$H^2(z)= (8 \pi G/3)\rho_{{\rm m}}=H^2_0 \, {\hat\Omega_{{\rm m}}}(z)$.
Therefore $H(z) = (H_0/3) \left[\left(\tilde{\zeta}_0 -3\right)(1+z)^{3/2}  - 
\tilde{\zeta}_0 \right]$.

		\paragraph{Cosmological tests}
We constrain the possible values of the bulk viscosity $\tilde{\zeta}_0$ 
using the cosmological tests of the Gold 2006 SNe Ia data sample 
\cite{Riess2006} composed by 182 SNe Ia, the Cosmic Microwave Background 
(CMB) shift parameter $R$ from the three-year WMAP observations, the Baryon 
Acoustic Oscillation (BAO) peak $A$ from the Sloan Digital Sky Survey (SDSS) 
and the Second Law of Thermodynamics (SLT).

For the SNe Ia we define the observational luminosity distance 
\cite{TurnerRiess2002,Riess2004}  in a flat cosmology and with the bulk viscous
matter component as $d_L = c(1+z)H^{-1}_0 \int_0^z E(z')^{-1} \; dz'$, where  $E(z) \equiv H(z)/H_0$.
The \emph{theoretical distance moduli} for the $i$-th supernovae with redshift $z_i$ is 
$ \mu^{{\rm t}}(z_i)=5\log_{10} [d_L(z_i)/1 {\rm Mpc}] +25 \;$, 
with $c$ the speed of light. The $\chi^2_{{\rm SNe}}$ statistical 
function becomes $\chi^2_{{\rm SNe}} (\tilde{\zeta}_0, H_0)  \equiv \sum_{k = 1}^{182}
   \left[\mu^{{\rm t}} (z_k , \tilde{\zeta}_0, H_0) - \mu_k \right]^2 / \sigma_k^2$, 
where $\mu_k$ is the \emph{observed} distance moduli for the $k$-th supernovae and 
$\sigma_k^2$ is the variance of the measurement.

The CMB shift parameter $R$  is defined as
$R \equiv \sqrt{\Omega^0_{{\rm m}}} \int^{Z_{{\rm CMB}}}_0 E(z')^{-1}\,dz'$, 
where $Z_{{\rm CMB}}=1089$ is the redshift of recombination
\cite{Melchiorri2003,WangMukherjee2006}. 
The observed value of the shift parameter $R$ is reported 
to be $R_{{\rm obs}}=1.70 \pm 0.03$ \cite{WangMukherjee2006}.

From the LSS data, we use the baryon acoustic oscillation 
(BAO) peak $A$ defined as  
$A \equiv \sqrt{\Omega^0_{{\rm m}}} E(z_1)^{-1/3} \left[ z^{-1}_1  \int^{z_{1}}_0 E(z')^{-1}\, dz' \right]^{2/3}$, 
where $z_1=0.35$ \cite{Eisenstein2005}. 
The observed value of SDSS-BAO peak is 
$A_{{\rm obs}}=0.469 \pm 0.017$ \cite{Eisenstein2005}.

We construct the joint $\chi^2$ function as $\chi^2_{{\rm total}} 
= \chi^2_{{\rm SNe}} + \chi^2_{{\rm CMB}} + \chi^2_{{\rm BAO}}$,  
where  $\chi^2_{{\rm SNe}}$ was defined above, and 
$\chi^2_{{\rm CMB}} = [(R -R_{{\rm obs}})/\sigma_R]^2 \;$ and $ \; 
\chi^2_{{\rm BAO}} = [(A -A_{{\rm obs}})/\sigma_A]^2$.

The generation of  \textit{local} entropy in the space--time is defined as 
$ T \, \nabla_{\mu} s^{\mu} = \zeta \nabla_{\mu} u^{\mu} = 3H\zeta$
\cite{Misner1973}, where $T$ is the temperature, $\nabla_{\mu} s^{\mu} $ 
is the rate at which entropy is being generated in a 
unit volume and $\zeta$ is the bulk viscosity coefficient.
The second law of thermodynamics can be written as 
$ T\nabla_{\mu} s^{\mu} \geq 0 \; \Rightarrow \; 3H\zeta \geq 0$.
Since $H$ is positive in an expanding universe then $\zeta$ has 
to be positive in order to preserve the validity of the 
second law of thermodynamics. Thus, in the present model 
a necessary condition is $\zeta_{{\rm m}} \geq 0$.

		\paragraph{Constraining the bulk viscous cosmological model}
We compute the  \textit{best estimated values} and 
``the goodness-of-fit'' of $\tilde{\zeta}_0$ and $H_0$ 
to the data through $\chi^2$-minimization, using SNe alone, 
and also considering the CMB and BAO. Then we compute the confidence 
intervals for $(\tilde{\zeta}_0,H_0)$ to constrain their possible values.
We obtain as best estimates using only SNe Ia: 
$(\tilde{\zeta}_0=1.71, H_0=62.30)$ with a $\chi^2_{\rm{min}}=161.2 \; (\chi^2_{\rm{d.o.f.}}=0.895)$, where hereinafter $H_0$ is in units of $\; {\rm km} \cdot {\rm sec}^{-1} \cdot {\rm  Mpc^{-1}}$.
And using SNe Ia, CMB and BAO together: $(\tilde{\zeta}_0=-0.30, H_0=52.97)$  with a 
$\chi^2_{\rm{min}}=400.5 \; (\chi^2_{\rm{d.o.f.}}=2.225)$.

\begin{center}
\begin{figure}
  \includegraphics[width=12cm]{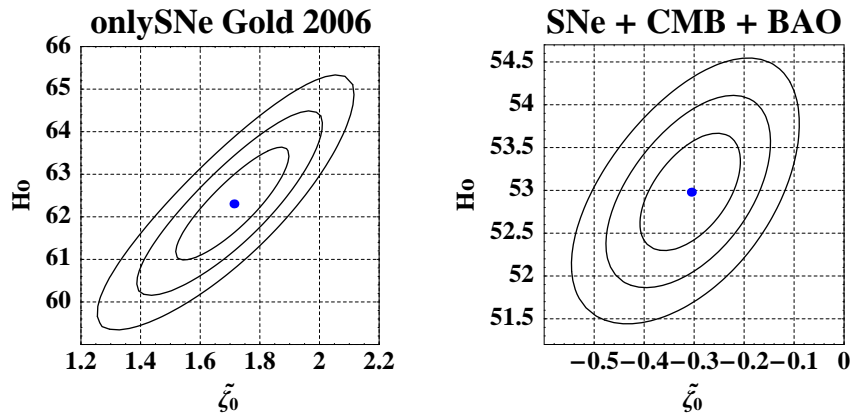}
  \caption{Joint confidence intervals for the parameters ($\tilde{\zeta}_0, H_0$) 
  for a bulk viscous matter-dominated universe, spatially flat and without any 
  dark energy component. We show the confidence 
  intervals of $68.3\% (1\sigma), \; 95.4 \% (3 \sigma)$ and $99.73 \% (5\sigma)$.
  The constraints are derived from the Gold 2006 SNe Ia data sample
  alone (left), and the joint Gold 2006 + CMB + BAO cosmological tests (right). }
  \label{ConfInterViscousMatter_ziHo}
\end{figure}
\end{center}

It can be seen that when we consider only SNe Ia we obtain  $\tilde{\zeta}_0 \geq 0$ 
with a 99.7\% confidence level and a reasonable value for the Hubble constant 
in the interval $ 59.5 \geq H_0 \geq 65.4$ as well as for $\chi^2_{\rm{d.o.f.}}=0.895$. 
However, when the same analysis using the three cosmological tests together 
(SNe Ia, CMB and BAO) is performed, we obtain negative values of $\tilde{\zeta}_0$ with at least 
99.7 \% confidence level that disagree with the SLT, a not so good estimation for 
$H_0$ of $ 51.4 \geq H_0 \geq 54.6$, and a bad  $\chi^2_{\rm{d.o.f.}}=2.225$. 
These results are illustrated in Figure \ref{ConfInterViscousMatter_ziHo}.

Therefore, it is possible to conclude that this model is not a 
good candidate for explaining the observed accelerated expansion 
of the universe as suggested in 
\cite{Fabris2006,Colistete2007,XinHe2006a}.

		\paragraph{Acknowledgments}
     This work is partly supported by grants 
     CIC-UMSNH 4.8, 4.9, PROMEP UMSNH-CA-22, SNI-20733 and COECYT CB0702137-2.

		\paragraph*{}

\end{document}